\documentclass[a4paper,10pt,twocolumn]{revtex4-1}

\usepackage[margin=25mm]{geometry}

\usepackage{graphicx}
\usepackage{mathrsfs}
\usepackage{xfrac}
\usepackage{amsmath} 
\usepackage{braket}
\usepackage{color}
\usepackage{placeins}
\usepackage{hyperref}
\usepackage{tabu}
\usepackage{colortbl}
\usepackage[dvipsnames]{xcolor}
\usepackage{soul}
\usepackage{algorithmic}
\usepackage{ulem}
\usepackage{appendix}

\newcommand{\be}{ \begin{equation} }
\newcommand{\ee}{ \end{equation} }

\newcommand{\bqa}{ \begin{eqnarray} }
\newcommand{\eqa}{ \end{eqnarray} }

\newcommand{\gdm}{\Gamma_\downarrow^{(m)}}
\newcommand{\gum}{\Gamma_\uparrow^{(m)}}

\newcommand{\gd}{\Gamma_\downarrow}
\newcommand{\gu}{\Gamma_\uparrow}
\newcommand{\gud}{\Gamma_\updownarrow}

\newcommand{\gdz}{\Gamma_\downarrow^{[0,0]}}
\newcommand{\guz}{\Gamma_\uparrow^{[0,0]}}
\newcommand{\gudz}{\Gamma_\updownarrow^{[0,0]}}

\newcommand{\li}{\mathrm{Li}}

\DeclareMathOperator{\dd}{d\!}

\newcommand{\trads}{\,\mathrm{Trad/s}}

\newcommand{\ignore}[1]{}

\begin{document}
	
\title{
Decondensation in non-equilibrium photonic condensates: when less is more}

\author{Henry J. Hesten}
\author{Robert A. Nyman}
\email{r.nyman@imperial.ac.uk}
\author{Florian Mintert}

\affiliation{Quantum Optics and Laser Science group,
	Blackett Laboratory, Imperial College London, Prince Consort Road, SW7
	2BW, United Kingdom}

\date{\today}

\begin{abstract}
	We investigate the steady state of a system of photons in a pumped dye-filled microcavity.
	By varying pump and thermalization the system can be tuned between Bose-Einstein condensation, multimode condensation, and lasing.
	We present a rich non-equilibrium phase diagram which exhibits transitions between these phases, including
	decondensation of individual modes under conditions that would typically favor condensation.
\end{abstract}  

\maketitle

Phase transitions in systems governed by quantum statistics at thermal equilibrium have been investigated intensely.
Canonical examples of these transitions are the formation of Cooper pairs in the superconductivity transition~\cite{Bardeen1957}, or the transition of a thermal cloud of bosonic atoms to a Bose-Einstein (BEC) condensate as temperature decreases~\cite{Anderson1995,Davis1995}.
The study of non-equilibrium phase transitions has spanned many disciplines from {\it classical} physics to social sciences~\cite{Lubeck2004}, and explained many phenomena, such as trends in society~\cite{Weidlich1991}, or traffic jams~\cite{Kerner1997}.
Only recently has attention turned to non-equilibrium phase transitions in {\it quantum} systems.
For example the condensation of polaritons in semiconductor~\cite{Kasprzak2006,Deng2006,Balili2007} or organic~\cite{Plumhof2014,Daskalakis2014} solids
has been observed outside thermal equilibrium.
Driven-dissipative many-particle quantum systems showing intricate phase diagrams \cite{PhysRevA.91.033609,PhysRevX.7.011016},
and bistability phenomena giving rise to coexisting phases have also been observed \cite{Sarchi2008,PhysRevX.7.011012}.

Our system of study is a gas of photons confined in a dye-filled microcavity which, when the dye is pumped, can be made to thermalize and Bose-Einstein condense~\cite{Klaers2010,Klaers2010a},
as predicted by thermal-equilibrium theory. Thermalization results from absorption and re-emission of light from the cavity by the dye~\cite{Schmitt2015}, 
which is limited by emission from the cavity~\cite{Marelic2015,Keeling2016}.
Thermal equilibrium is thus always imperfect and breaks down completely if the cavity is far detuned from the dye molecular resonance~\cite{Kirton2013, Kirton2015}. The system then features multimode condensation~\cite{Marelic2016,Keeling2016} as a clear signature of non-equilibrium behavior.
It has been noted that multimode systems driven far from equilibrium can 
show multimode condensation~\cite{vorberg2013}, and that the kinetics of 
two-mode laser systems can be made to show a sort of minimalist 
Bose-Einstein condensation~\cite{redlich2016, leymann2017}.

Extrapolating from prior experiments on non-equilibrium phase transitions~\cite{PhysRevX.7.011016,PhysRevX.7.011012}, one would expect that condensation is always favored by an increase in the pump rate~\cite{Kasprzak2006,Klaers2010} and a decrease in thermalization rate.
This is typically true for the overall fraction of photons in condensed modes.
For individual modes however, 
we also find the opposite behavior in the regime far out of equilibrium.
That is, modes with condensed photons can loose their macroscopic occupation as the pump rate is increased, which is similar to decreasing entropy as temperature increases, or negative heat capacities \cite{michaelian2007}.
This gives rise to a highly complex dependence of the steady state on properties like pump rate, geometry, and the time-scale of thermalization.
Based on a microscopic model~\cite{Kirton2013} we predict the dependence of multimode condensation on such properties.
Despite the non-linear nature of the system we find analytic laws characterizing  condensation and decondensation that coincide accurately with numerically exact solutions. 

In dye-filled microcavities only one longitudinal mode of the harmonic cavity is sufficiently close to resonance with the dye molecules;
the dynamics can then be reduced to a two-dimensional model with transverse modes labeled by $m=[m_x,m_y]$.
Taking into account loss through the mirrors and spontaneous emission from dye molecules into modes that are not confined within the cavity results in the well-established equation of motion \cite{Keeling2016}
\be\begin{split}
  \frac{\dd n_m}{\dd t}=-\kappa n_m+&
  \rho\Gamma_\downarrow^{(m)} f_m (n_m+1) \\
  +&\rho\Gamma_\uparrow^{(m)}(f_m-1) n_m\ ,
  \label{eq:dndt}
\end{split}
\ee
for the average occupation $n_m$ of mode $m$,
where $\kappa$ is the cavity decay constant, $\gum$ and $\gdm$ are respectively the rates of absorption from and emission into mode $m$, and $\rho$ is the areal density of molecules;
$f_m$ is the fraction of excited molecules interacting with mode $m$,
and it is given in terms of the fraction $f(\mathbf{r})$ of molecules at point ${\bf r}$ which are excited
and the mode-profile
$\psi_m(\mathbf{r})$ via the relation ${f_m=\int d^2{\bf r} f(\mathbf{r}) |\psi_m(\mathbf{r})|^2}$.
The dynamics of the excited-state population is governed by
\begin{equation}
  \label{eq:dfdt}
  \frac{\partial f(\mathbf{r})}{\partial t} = -\Gamma^{tot}_{\downarrow}(\mathbf{r})f(\mathbf{r}) + \Gamma^{tot}_{\uparrow}(\mathbf{r})(1-f(\mathbf{r}))\ ,
\end{equation}
in terms of the rates of total absorption and emission, $\Gamma^{tot}_{\downarrow}(\mathbf{r})$ and $\Gamma^{tot}_{\uparrow}(\mathbf{r})$,
which depend on the mode occupations via
\begin{equation}
  \label{eq:gtot}
  \Gamma^{tot}_{k}(\mathbf{r}) = \Gamma_{k}(\mathbf{r}) + \sum_m  |\psi_m(\mathbf{r})|^2 \Gamma_k^{(m)}(n_m+\delta_{k\downarrow})\ 
\end{equation}
with $k=\uparrow,\downarrow$ and $\delta_{\uparrow\downarrow}=1-\delta_{\downarrow\downarrow}=0$,
where $\Gamma_{\uparrow/\downarrow}(\mathbf{r})$ are the pump rate of molecules by the laser, and the decay of molecules not captured by emission into the modeled cavity modes.

As the thermalization process occurs when excitations are exchanged between cavity modes and dye molecules,
we compare the rate of absorption $\rho \gum$ to the rate of loss $\kappa$ and define the thermalization coefficient $\gamma = \rho \Gamma_{\uparrow}^{[0,0]}/\kappa$ \footnote{The thermalization coefficient $\gamma$ is directly related to the detuning, but since this relation depends on the absorption spectrum of the molecules there is no simple analytical form for it.
Nevertheless, since $\gamma$ is the more informative quantity, we will mostly refer to thermalization in the following.}.
We choose parameter values appropriate to real experiments,
but specify all values in units of cavity decay $\kappa$ (which for typical experiments is of the order of $10^{9}/$s) and harmonic oscillator length $L$ (the mean spatial extent of the lowest cavity mode).
The shape of absorption and emission profiles $\Gamma_{\uparrow/\downarrow}^{(m)}$ for the individual modes are extracted from experimental data \cite{nyman_robert_andrew_2017_569817} and
their peaks are set to $ 1.2 \times 10^{-9} \, \kappa$ (see appendix~\ref{supp:spectra}).
We consider a slightly anisotropic cavity with mode spacings $\omega_x=\omega_y/1.01 =3 \times 10^{4} \, \kappa$,
and a detuning between the molecular resonance frequency and the lowest cavity eigenfrequency ranging from $-3.5\times 10^{5}\kappa$ to $-1.89\times 10^{5}\kappa$.
The pump of the molecules of areal density $\rho = 10^{12}/L^2$ 
has a Gaussian profile with width $20L$,
and the decay rate $\Gamma_{_\downarrow}$ of excited molecules is set to $\Gamma_{_\downarrow} = \kappa \,/\, 4$.
The anisotropy of the cavity is chosen in order to avoid degeneracies, but it is sufficiently small so that mode pairs $[m_x,m_y]$ and $[m_y,m_x]$ behave almost identically, and their condensation thresholds are hardly distinguishable. We will therefore only discuss modes $[m_x,m_y]$ with $m_x\le m_y$.
In order to arrive at a finite-dimensional problem, we consider cavity modes with $m_x+m_y\le 6$ only.

\begin{figure}[t]
	\centering
	\includegraphics[width=\columnwidth]{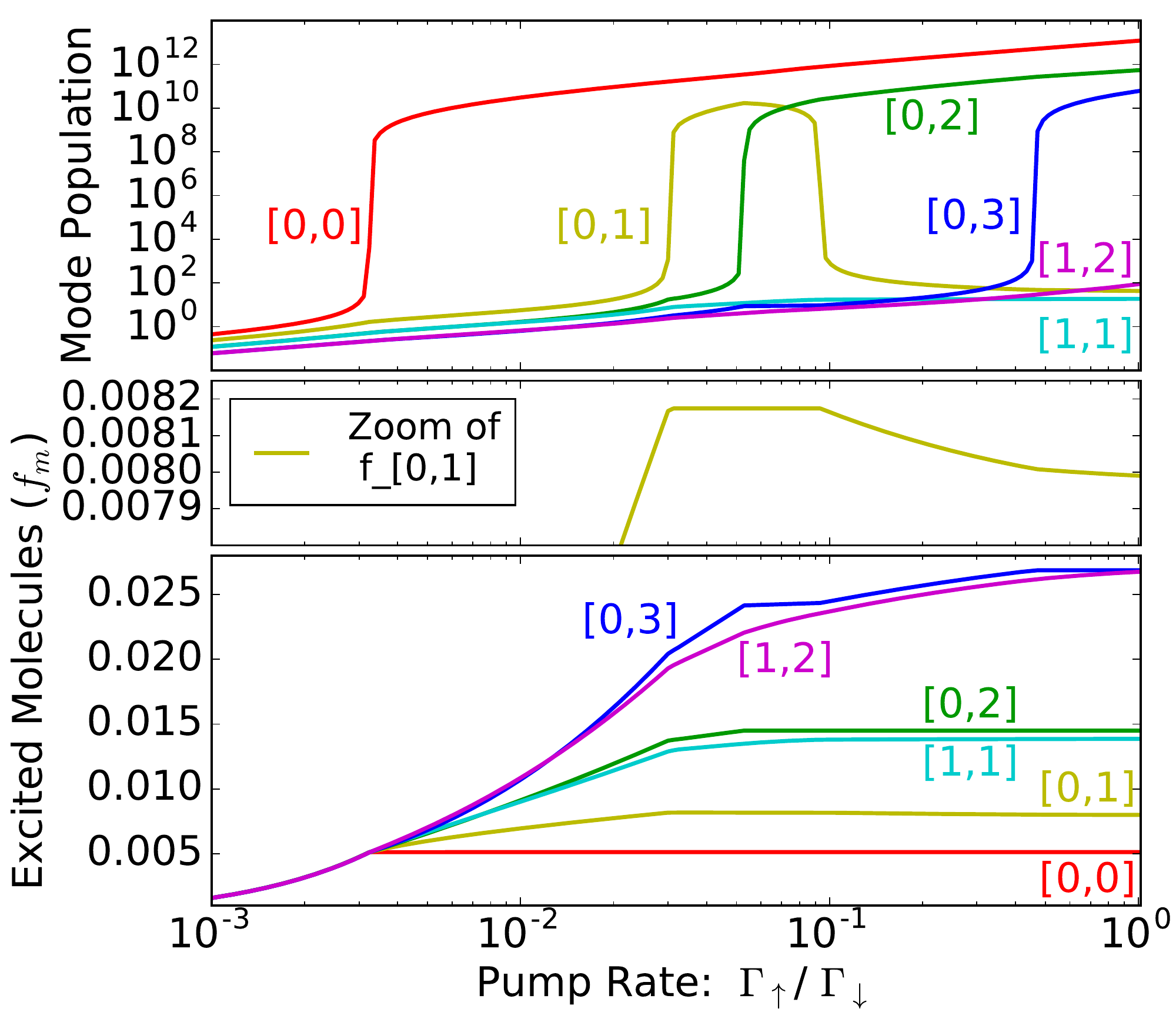}
	\caption{
		The upper panel depicts the mode populations $n_m$ as functions of pump rate for a constant thermalization rate $\gamma=1.8 $.
		One can clearly see how mode $[0,1]$ condenses and then de-condenses with increasing pump rate.
		The lower panel depicts the fraction of excited molecules $f_m$ accessible to any mode, and substantiates that the decondensation of mode $[0,1]$ is congruent with a decrease of $f_{[0,1]}$. This is highlighted in the central panel, which is a magnification of the lower panel.}
	\label{fig1}
\end{figure}
Fig.~\ref{fig1} (top panel) depicts the stationary solutions for mode occupations as functions of pump rate,
and one can see step-like increases and decreases of the populations, {\it i.e.} condensation and decondensation of individual modes at specific values of pump rate.
The BEC phase is defined by condensation in the lowest cavity mode only,
whereas a multimode condensate contains additional condensed modes.
Any phase with one or more condensed modes, but an un-condensed ground mode is considered a laser \footnote{The distinction we make here is that in 
Bose-Einstein and multimode condensate phases both absorption and 
emission play important roles, whereas (stimulated) emission dominates 
the behavior in the laser phase.}
and any phase without any condensed modes will be called un-condensed.
With increasing pump rate, Fig.~\ref{fig1} (top panel) thus features the transition from an un-condensed phase to BEC, followed by a transition from BEC to a multimode condensate, and three transitions between different multimode condensates.

\begin{small}
	\begin{figure*}[t]
		\centering
		\begin{minipage}{0.62\textwidth}
			\includegraphics[width=\columnwidth]{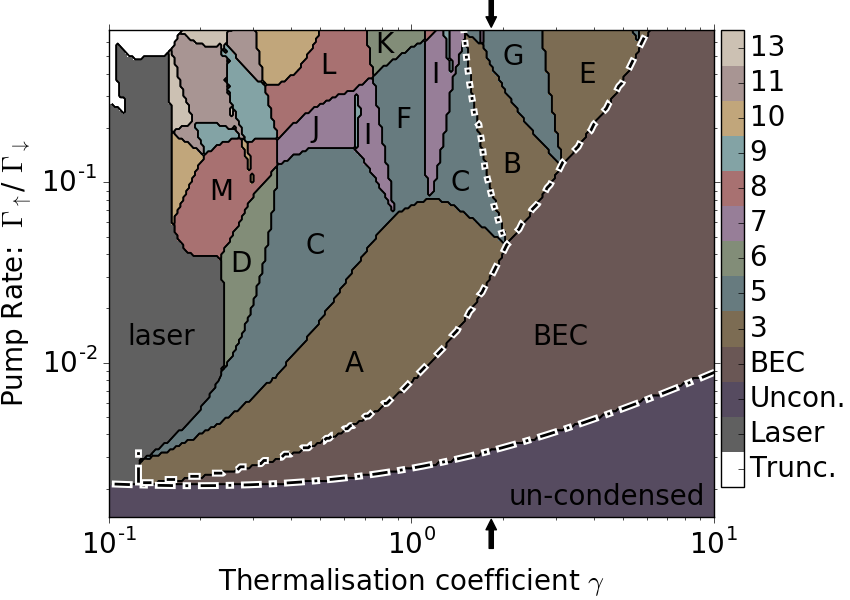}
		\end{minipage}
		\begin{minipage}{0.37\textwidth}
			{\renewcommand{\arraystretch}{1.2}
				\setlength\tabcolsep{2pt}
				\begin{tabular}{|c|cccccccc|}
					\hline
					& \multicolumn{7}{p{3.6cm}}{\hspace{1cm}Condensed modes} &\\\hline
					BEC &[0,0]&&&&&&&\\ \hline
					A &[0,0] & [0,1]&&&&&&\\ \hline
					B &[0,0] && [0,2] &&&&&\\ \hline
					C &[0,0] & [0,1] & [0,2]&&&&&\\ \hline
					D &[0,0] & [0,1] & [0,2] & [1,1]&&&&\\ \hline
					E &[0,0] &&&& [0,3] &&&\\ \hline
					F &[0,0] & [0,1] &&& [0,3]&&&\\ \hline
					G &[0,0] && [0,2] && [0,3]&&&\\ \hline
					I &[0,0] &[0,1]& [0,2] && [0,3] &&& \\ \hline
					J &[0,0] & [0,1] & [0,2] &&&[0,4]&&\\ \hline
					K &[0,0] & [0,1] & && [0,3] &&& [2,2] \\ \hline
					L &[0,0] & [0,1] & [0,2] & &&[0,4]&& [2,2] \\ \hline
					M &[0,0] & [0,1] & [0,2] & [1,1]&[0,3]&&&  \\ \hline
				\end{tabular}
			}
		\end{minipage}
		\caption{Phase diagram of the photon gas as function of pump rate $\Gamma_\uparrow$ and thermalization coefficient $\gamma$. 
			Different colors indicate the number of condensed modes with the sole exceptions of `laser', which indicates condensed phases of the system without condensation in the ground mode, and `truncated' indicating potential truncation errors as the highest considered mode has condensed.
			Capital letters indicate which modes are condensed in a given region;
			narrow areas in which only one of modes $[m_x,m_y]$ and $[m_y,m_x]$ are condensed are treated as if both modes are condensed in order to avoid too detailed structures; the fine structures between regions `I' and `J' and 
			adjacent to region `M' are a result of this.
			Rough phase boundaries in the top left region are due to limited numerical accuracy.
			The analytic estimates for various phase transitions
			indicated by white lines 
			coincide very accurately with the numerically obtained thresholds.
			The cut for $\gamma=1.8$ through the phase diagram depicted in Fig.~\ref{fig1} is indicated by bold arrows.
		}
		\label{fig2}
	\end{figure*}
\end{small}
Despite the system's complex, non-linear behavior, we can develop an understanding of (de)-condensation in terms of the decomposition of the total pump rate into individual contributions given in Eq.~\eqref{eq:gtot}.
For low populations $n_m$ of all modes, the total rates of absorption/emission can be well approximated by direct pumping and loss, {\it i.e.}
Eq.~\eqref{eq:gtot} reduces to
$\Gamma^{tot}_{\uparrow/\downarrow}(\mathbf{r}) \simeq \Gamma_{\uparrow/\downarrow}(\mathbf{r})$,
so that the pumping of dye molecules (Eq.~\eqref{eq:dfdt}) is proportional to the external pumping.
In the case of strongly occupied modes, on the other hand, the approximation
$\Gamma^{tot}_{\uparrow/\downarrow}(\mathbf{r}) \simeq \sum_m |\psi_m(\mathbf{r})|^2 \, \Gamma^{m}_{\uparrow/\downarrow}n_m$
in terms of the rates $\Gamma^{m}_{\uparrow/\downarrow}n_m$ of stimulated absorption and emission holds,
and the proportion of excited dye molecules is largely independent of external pumping but depends mostly on the interaction with the condensed mode or modes.
Molecules in spatial domains where $|\psi_m(\mathbf{r})|^2 \, \Gamma^{m}_{\uparrow/\downarrow}n_m\gg \Gamma_{\uparrow/\downarrow}(\mathbf{r})$ are
thus {\it clamped} to mode $m$.

As pumping is increased the excitation of clamped molecules decreases,
but the excitation of un-clamped molecules grows. In particular, un-condensed modes thus experience a growing reservoir of excitations that can help them to gain enough population to condense.
As a mode condenses it also starts to clamp dye molecules and the number of molecules clamped to this mode grows with increasing population $n_m$.
This mode thus enters a competition for access to excitations with the other modes, and potentially reduces the access for other modes, which can result in their decondensation.
This effect can be explicitly seen in the fraction of excited molecules $f_m$ accessible to mode $m$ as shown in Fig.~\ref{fig1} (center and bottom panel).
The center panel magnifies $f_{[0,1]}$ and shows that $f_{[0,1]}$ decreases with increasing pump rate after the condensation of mode $[0,2]$, despite the general (and expected) trend that $f_m$ grows with increasing pump rate.
Quite strikingly, the decrease in accessible excited dye molecules required for decondensation is rather minute and visible only in the magnified panel.
Nevertheless, it has an extremely significant impact and can result in decondensation of a mode.

With this qualitative understanding of decondensation at hand, we can now proceed to a quantitative prediction of condensation thresholds; a more detailed analysis and the explicit analytic solutions can be found in appendix~\ref{sec:analytic}.
The mode population $n_m$ for stationary solutions to Eq.~\eqref{eq:dndt} diverges if $f_m$ approaches the critical value
\begin{equation}
\begin{split}
f_m^{c} = \frac{\gum+\frac{\kappa}{\rho}}{\gdm +\gum}\ .
\label{eq:fmcrit}
\end{split}
\end{equation}
Neglecting contributions from uncondensed modes to the total pump rate (Eq.~\eqref{eq:gtot}),
one obtains the stationary solutions
\begin{equation}
  f_m^{(s)} = \int d^2{\bf r} |\psi_m({\bf r})|^2 \frac{ \Gamma^{tot}_{\uparrow}({\bf r})}{\Gamma^{tot}_{\uparrow}({\bf r}) + \Gamma^{tot}_{\downarrow}({\bf r})}
  \label{eq:fm}
\end{equation}
of Eq.~\eqref{eq:dfdt}
 for {\it all} modes
 as functions of the mode populations $n_m$ of the {\it condensed} modes.
Gain clamping implies the condition $f_m^{s}\lesssim f_m^{c}$ for the condensed modes, which, in turn determines $n_m$ for all condensed modes.
This permits identification of pump rates (or other system parameters) that achieve the condensation condition for an originally uncondensed mode.
The thresholds for Bose-Einstein condensation, multimode condensation and the decondensation of mode $[0,1]$ while mode $[0,2]$ is condensed are depicted as white lines in the phase diagram depicted in Fig.~\ref{fig2}.

The accuracy of these estimates can be verified by comparisons to numerically exact solutions.
The steady-state solution of Eq.~\eqref{eq:dndt}, together with Eq.~\eqref{eq:fm},
can be found with algebraic root finding routines,
and we verified explicitly that the obtained solutions coincide with the solutions obtained by propagation until a stationary state is reached.
Defining phase boundaries can be done unambiguously in the thermodynamic limit,
but it poses an intricate problem in systems of finite size.
Since numerically exact solutions do not result in diverging mode populations,
we employ large increases in the population of a single mode under small changes of driving conditions (as depicted in Fig.~\ref{fig1} (top panel)) as indicators of condensation threshold~\cite{Kasprzak2006,Marelic2016}.
Quantitatively, we define the threshold of condensation (decondensation) as an increase (decrease) in population greater than 3 orders of magnitude over an increase in pump rate of 10\%,
but due to the sharp nature of all observed transitions, the identification of threshold is largely independent of the explicitly chosen numerical values, see appendix~\ref{supp:threshold}.

As depicted in Fig.~\ref{fig2} the system at lower pump rates is in the un-condensed phase, with no condensed modes.
At large thermalization coefficient and pump rate the system is in the BEC phase.
As the thermalization coefficient is decreased from here,
the system passes through various multimode phases, until the ground mode is no longer condensed.
Under these conditions the system behaves like a laser, where condensation is a consequence of stimulated emission, rather than a condensate, for which absorption and emission play equally important roles.
The trend that strong pumping and weak thermalization favors condensation is observed for the onset of Bose-Einstein and multi-mode condensation, but in the regime of multi-mode condensation the behavior becomes more complicated and decondensation can be induced through an increase in pump rate $\Gamma_\uparrow$ or a decrease in thermalization coefficient $\gamma$.
Given the extremely sensitive dependence of the system's phase on the excitation of dye molecules,
there are several instances where small changes in pumping or thermalization result in a substantial redistribution of populations between the modes, and multiple triple- and quadruple-points.

Eventually, a critical assessment of experimental relevance is in order.
Current experiments are performed for $\gamma$ ranging between $0.2$ and $5$ and pump rates reach rates ten times the threshold value~\cite{Marelic2016}.
Exploring the upper half of the phase diagram presented in Fig.~\ref{fig2}, requires substantially stronger pumping than currently realized.
Since pulsed lasers can achieve peak pump rates three orders of magnitude higher than CW lasers, with pulse durations longer than the time taken to reach steady state, this seems a perfectly viable option.
The model we have used is applicable to a wide range of systems which need only satisfy a few criteria: an optical environment with a well-defined ground state, a fluorescent gain medium, and re-scattering of light faster than loss resulting in thermalization.
We thus expect that 
multimode condensation and decondensation both inside and outside the lasing regime will be observable in plasmonic lattices coupled to dyes, which have recently shown condensation~\cite{hakala2017}, semiconductors in photonic crystal resonators~\cite{deLeeuw16}, and also more conventional laser systems.
Exciton-polariton condensates (both semi-conductor and organic), however, differ from photon condensates in that the mixed light-matter excitations relax directly through 
their matter component, whereas in photon and plasmon condensates 
equilibration occurs via exchanges between weakly-coupled light and 
molecular excitations. Our predictions therefore do not translate directly, but it would be intriguing to identify similar mechanisms in these condensates.

The rich interplay between the different modes observed in Fig.~\ref{fig2} also offers great opportunities for the creation of tailored states of light \cite{hisch2013},
since the mode structure can easily be influenced through the shape of the cavity \cite{flatten2015}, lattice~\cite{hakala2017} or crystal structure~\cite{deLeeuw16}, and the spatial pump profile can be varied.
Since fluctuations are most relevant near phase transitions, we expect this rich phase diagram to be a fruitful tool in the search for unusual quantum correlations.

We acknowledge fruitful discussions with David Newman, Peter Kirton and Jonathan Keeling,
and financial support from the UK~EPSRC (via fellowship EP/J017027/1 and the Controlled Quantum Dynamics CDT EP/L016524/1), and the ERC (via ODYCQUENT grant).
The data underlying this article are available at \cite{nyman_robert_andrew_2017_569817}.

\appendix
	
	\vspace{1cm}
	{\LARGE \bf Appendices}	
	\vspace{0.3cm}

In appendix~\ref{sec:analytic} we provide a detailed, analytic derivation of phase boundaries resulting from gain clamping. Appendix~\ref{supp:spectra} contains a discussion of the absorption and emission spectra used in the simulation, followed by appendix~\ref{supp:threshold} with a justification for the choice of threshold condition. In appendix~\ref{supp:size}, we show how a variation of pump profile induces multimode condensation and decondensation, similarly to the variation of thermalization that is discussed in the main paper.
	
\section{Analytic Phase Boundaries}
\label{sec:analytic}

As sketched in the main paper, phase boundaries can be estimated analytically with high accuracy based on the clamping mechanism.
Here we discuss the derivation of the boundaries depicted in Fig.~\ref{fig2} in more detail.
In the presently considered case of spatially homogeneous pumping,
both pump profile $\gu({\bf r})$ and molecular decay $\gd({\bf r})$ are constant and will be denoted by $\gu$ and $\gd$ in the following.
We will also assume a rotationally invariant cavity with degenerate modes.

The population of mode $m$ in the stationary solution of Eq.~(1) reads
\be
n_m=\frac{\Gamma_\downarrow^{(m)} f_m}{\frac{\kappa}{\rho}-\Gamma_\downarrow^{(m)} f_m-\Gamma_\uparrow^{(m)}(f_m-1)}\ .
\label{supp_nm}
\ee
It diverges if $f_m$ reaches the critical density
\be
f_m^{(c)}=\frac{\Gamma_\uparrow^{(m)}+\frac{\kappa}{\rho}}{\Gamma_\uparrow^{(m)}+\Gamma_\downarrow^{(m)}}
\label{supp:fc}
\ee
of excited molecules.
Approximating Eq.~\eqref{eq:gtot} according to a given phase transition, and equating $f_m^{(c)}$ with Eq.~\eqref{eq:fm}, thus allows us to estimate phase boundaries.

\subsection{Boundary between the un-condensed phase and single-mode condensation}
\label{sec:firstthreshold}

The decay of the molecular excited states is dominated by spontaneous emission into free space at a rate $\Gamma_\downarrow$, which is much larger than the spontaneous emission rate into any given mode $\Gamma_\downarrow^{(m)}$ because the effective Purcell enhancement of spontaneous emission into any given cavity mode is weak.
As the mode frequencies are below the molecular resonance we can approximate the critical molecular excitation as $f_m^{(c)} \gtrsim \gum/\gdm$.
Since, in addition, the molecules are not saturated, the molecular excitation can be approximated as $f_m \approx \gu/\gd$.
Near threshold the molecular excitation must be close to the critical value, {\it i.e.} $f_m \approx f_m^{(c)}$, so that
the inequality $\gu\gdm\gtrsim \gd\gum$
follows.
From the established condition $\gd \gg \gdm$, 
one can thus conclude,
that external pumping is far stronger than re-absorption of light from cavity modes, {\it i.e.} $\gu \gg \gum$.
Below condensation threshold, where no mode population $n_m$ is macroscopically large, Eq.~\eqref{eq:gtot} can therefore be approximated as
\begin{equation}
\begin{split}
\Gamma^{tot}_{\downarrow}(\mathbf{r}) &\simeq \Gamma_{\downarrow}\ ,\\
\Gamma^{tot}_{\uparrow}(\mathbf{r}) &\simeq \Gamma_{\uparrow}\ .
\label{eq:Eq3firstthreshold}
\end{split}
\end{equation}
Within this approximation, Eq.~\eqref{eq:fm} reduces to
\begin{equation}
f_m^{(s)} = \int d^2{\bf r} |\psi_m({\bf r})|^2 \frac{ \Gamma_{\uparrow}}{\Gamma_{\uparrow} + \Gamma_{\downarrow}}=\frac{ \Gamma_{\uparrow}}{\Gamma_{\uparrow} + \Gamma_{\downarrow}}\ .
\end{equation}
Equating this with $f_m^{(c)}$ (Eq.~\eqref{supp:fc}) yields the threshold pump rate
\be
\frac{\Gamma_{\uparrow}}{\Gamma_{\downarrow}}=\frac{\Gamma_\uparrow^{(m)}+\frac{\kappa}{\rho}}{\Gamma_\downarrow^{(m)}-\frac{\kappa}{\rho}}\ .
\label{eq:thresh1}
\ee

Given the assumption of sufficiently small mode populations $n_m$, Eq.~\eqref{eq:thresh1} applies to the first condensation threshold only,
{\it i.e.} it predicts the condensation of the mode $m$ with the lowest threshold pump rate.
If this is the lowest mode $[0,0]$, then Eq.~\eqref{eq:thresh1} describes the threshold of Bose-Einstein condensation; otherwise it predicts the onset of lasing.

Condensation into higher modes is more favorable than condensation into the ground mode, because the higher modes are closer to molecular resonance and therefore couple more strongly to the molecules. Thermalization of photons is therefore essential for the BEC phase as this redistributes photons from higher modes into the ground mode. If the system is condensed into a laser instead of a BEC, it is likely to remain a laser as pump rate increases, since the thermalization is unable to overcome the clamping of the molecules. This permits the estimation of the boundary between the laser and BEC phases for low pump powers, by considering the mode with the lowest threshold pump rate.

\subsection{Boundary between single-mode and multi-mode condensation}
\label{sec:secondthreshold}

The boundary between  single-mode and multi-mode condensation can be derived very analogously to the discussion in Sec.~\ref{sec:firstthreshold}.
Since, however, the population of the ground state mode is macroscopic, Eq.~\eqref{eq:gtot} can no longer be approximated by Eq.~\eqref{eq:Eq3firstthreshold}.
Instead, it should be replaced by
\begin{equation}
\begin{split}
\Gamma^{tot}_{\uparrow}(\mathbf{r}) &\simeq \Gamma_{\uparrow}(\mathbf{r}) +|\psi_{[0,0]}(\mathbf{r})|^2 \Gamma_\uparrow^{[0,0]}n_{[0,0]}\\
\Gamma^{tot}_{\downarrow}(\mathbf{r}) &\simeq \Gamma_{\downarrow}(\mathbf{r}) + |\psi_{[0,0]}(\mathbf{r})|^2 \Gamma_\downarrow^{[0,0]}n_{[0,0]}\ .
\label{eq:Eq3secondthreshold}
\end{split}
\end{equation}
such that, Eq.~\eqref{eq:fm} reduces to
\begin{equation}
f_m^{(s)} = \int d^2{\bf r} |\psi_m({\bf r})|^2 \frac{\Gamma_{\uparrow}({\bf r})+|\psi_{[0,0]}(\mathbf{r})|^2 \Gamma_\uparrow^{[0,0]}n_{[0,0]}}{\Gamma_{\updownarrow}({\bf r})+|\psi_{[0,0]}(\mathbf{r})|^2\Gamma_\updownarrow^{[0,0]}n_{[0,0]}}\ ,
\label{supp:fmssecondthreshold}
\end{equation}
with the short hand notations
$\Gamma_\updownarrow=\gd+\gu$ and
$\Gamma_\updownarrow^{[0,0]}=\guz+\gdz$.

Determining $f_m^{(s)}$ requires knowledge of the ground-state population $n_{[0,0]}$.
The actual value of $f_{[0,0]}^{(s)}$ is close to its critical value, but does not exactly coincide with it,
since $n_{[0,0]}$ is macroscopically large, but strictly speaking not diverging.
Since Eq.~\eqref{supp_nm} diverges for $f_{[0,0]}^{(s)}\to f_{[0,0]}^{(c)}$, it is not a good starting point for an estimate of $n_{[0,0]}$ for $f_{[0,0]}^{(s)}\lesssim f_{[0,0]}^{(c)}$.
Eq.~\eqref{supp:fmssecondthreshold}, on the other hand, can be solved for a finite value of $n_{[0,0]}$ by equating $f_{[0,0]}^{(s)}$ with $f_{[0,0]}^{(c)}$ as given in Eq.~\eqref{supp:fc}.
The population of the ground mode below the second condensation threshold can thus be approximated by the relation
\be
f_{[0,0]}^{(s)}=\frac{\Gamma_\uparrow^{[0,0]}+\frac{\kappa}{\rho}}{\Gamma_\uparrow^{[0,0]}+\Gamma_\downarrow^{[0,0]}}\ ,
\ee 
with $f_{[0,0]}^{(s)}$ defined in Eq.~\eqref{supp:fmssecondthreshold}.
The corresponding value of $n_{[0,0]}$ then allows the calculation of the second condensation threshold by equating $f_m^{(s)}$ (for $m\neq [0,0]$) with $f^{(c)}_m$.
The mode that achieves this threshold with the the lowest pump rate is then the mode that condenses in addition to the ground mode.

Evaluating these thresholds requires the explicit forms
\begin{equation}
\begin{split}
\left|\psi_{[0,0]}\right|^2 =& \frac{1}{\pi} \frac{1}{L^2} \; e^{\sfrac{-(x^2+y^2)^2}{L^2}} \\
\left|\psi_{[0,1]}\right|^2 =& 2\left|\psi_{[0,0]}\right|^2 \frac{x^2}{L^2} \\
\left|\psi_{[0,2]}\right|^2 =& \frac{1}{2} \left|\psi_{[0,0]}\right|^2 \left[4 \left(\frac{x}{L}\right)^4 - 4\left(\frac{x}{L}\right)^2 + 1\right]\\
\left|\psi_{[0,3]}\right|^2 =& \frac{1}{3} \left|\psi_{[0,0]}\right|^2 \left[4 \left(\frac{x}{L}\right)^6 - 12\left(\frac{x}{L}\right)^4 + 9\left(\frac{x}{L}\right)^2\right]\nonumber
\end{split}
\end{equation}
of the mode functions for a two-dimensional harmonic oscillator.
The integrals can be evaluated analytically and read
\begin{equation}
\begin{split}
f_{[0,0]} =& 2 I_0 \\
f_{[0,1]} =& 2 I_2 \\
f_{[0,2]} =& \frac{3}{2}I_4 - 2I_2 +  I_0  \\
f_{[0,3]} =& \frac{5}{6}I_6 - 3I_4 + 3 I_2\ , \nonumber
\end{split}
\end{equation}
with
\begin{equation}
I_k=\frac{\gu}{\gud}\int \; \dd r \; \frac{1+\sigma\;  e^{-r^2}}{1+\eta\; e^{-r^2}} \; r^{(k+1)} \, e^{-r^2}\ ,\nonumber
\end{equation}
and
\begin{equation}
\begin{split}
\sigma=&\frac{\guz}{\gu}\frac{n_{[0,0]}}{\pi} \ ,\\
\eta=&\frac{\gudz}{\gud}\frac{n_{[0,0]}}{\pi}\ .\nonumber
\end{split}
\end{equation}
The $I_k$ can be expressed explicitly
\begin{equation}
\begin{split}
I_0 =& \frac{\gu}{2\eta\gud}\left(\sigma+ \left(1-\frac{\sigma}{\eta} \right)\log(1+\eta)\right) \\
I_2 =& \frac{\gu}{2\eta\gud}\left(\sigma\left(1-\frac{\pi^2}{6\eta}\right)
-\li_2(-\eta) +\right.\\
&\left.\frac{\sigma}{\eta}\left(\frac{\log(1+\eta)}{2}\log\left(\frac{1+\eta}{\eta^2}\right) + 
\li_2\left(\frac{1}{1+\eta}\right)\right)\right)
\label{eq:f1} \\
I_4 =& \frac{\gu}{\eta\gud}\left(\sigma - \frac{1-\sigma}{\eta} \li_3(-\eta)\right) \\
I_6 =& \frac{3\gu}{\eta\gud}\left(\sigma - \frac{1-\sigma}{\eta} \li_4(-\eta)\right)
\nonumber
\end{split}
\end{equation}
in terms of the polylogarithm functions
\begin{equation}
\li_j(x)=\sum_{i=1}^{\infty}\frac{x^i}{i^j}\ .\nonumber
\end{equation}
The condition $f_m^{(s)}=f_m^{(c)}$ then determines the pump rate $\gu$ at which multi-mode condensation occurs.
This condition can not be solved for $\gu$ analytically, but we found it to have a unique numerical solution.

With the principles of these analytic estimates,
one can also explain which mode condenses at the lowest rate.
The thermalization process redistributes photons from higher modes into lower modes, thus making it difficult for the higher mode to condense.
At low thermalization rate mode $[0,1]$ condenses at the lowest pump rate to form phase `A' in Fig.~\ref{fig2}. As thermalization increases, molecules overlapping mode $[0,1]$ become clamped by the ground state, and any increase in the population of mode $[0,1]$ is transferred to the ground state. The thermalization between mode $[0,2]$ and $[0,0]$ is less than between $[0,1]$ and $[0,0]$, therefore increasing the thermalization coefficient has a smaller effect on $[0,2]$ than $[0,1]$. This allows mode $[0,2]$ to condense at a lower pump rate than $[0,1]$ for large thermalizations.
This explains why the three-mode phase changes from `A' to `B' to `E' as thermalization increases, and why a larger pump rate is required. 

\subsection{Decondensation}
\label{sec:thirdthreshold}

Last, but not least, let us discuss the decondensation of mode $[0,1]$ while modes $[0,0]$ and $[0,2]$ are condensed, {\it i.e.} the transition from `C' to `B' in Fig.~\ref{fig2}.

Above the decondensation threshold (`B'), modes $[0,0]$, $[0,2]$, and $[2,0]$ are condensed. We therefore replace Eq.~\eqref{eq:Eq3secondthreshold} with
\begin{equation}
\begin{split}
\Gamma^{tot}_{\uparrow}(\mathbf{r}) &\simeq \Gamma_{\uparrow}(\mathbf{r}) + \sum_{p\in S_{0,2}} |\psi_p(\mathbf{r})|^2 \Gamma_\uparrow^{(p)} n_p\\
\Gamma^{tot}_{\downarrow}(\mathbf{r}) &\simeq \Gamma_{\downarrow}(\mathbf{r}) + \sum_{p\in S_{0,2}} |\psi_{p}(\mathbf{r})|^2 \Gamma_\downarrow^{(p)}n_{p}\ ,
\label{eq:Eq3thirdthreshold}
\end{split}
\end{equation}
where $S_{0,2} = \{\,[0,0],\,[0,2],\,[2,0]\,\}$.
Eq.~\eqref{eq:fm} therefore reduces to
\begin{equation}
f_m^{(s)} = \int d^2{\bf r} |\psi_m({\bf r})|^2 \frac{\Gamma_{\uparrow}({\bf r})+\sum|\psi_{p}(\mathbf{r})|^2 \Gamma_\uparrow^{p}n_{p}}{\Gamma_{\updownarrow}({\bf r})+\sum|\psi_{p}(\mathbf{r})|^2\Gamma_\updownarrow^{p}n_{p}}\ ,
\label{supp:fmsthirdthreshold}
\end{equation}
with the summations performed over $S_{0,2}$.
As modes $[0,0]$, $[0,2]$ and $[2,0]$ are condensed their populations are determined from the condition $f^{(s)}_{m} = f^{(c)}_{m}$.
This permits evaluation of $f_{[0,1]}^{(s)}$ following Eq.~\eqref{supp:fmsthirdthreshold}.
Unlike the case of multi-mode condensation discussed above in Sec.~\ref{sec:secondthreshold}, the integrals can no longer be evaluated analytically, but Eq.~\eqref{supp:fmsthirdthreshold} can readily be integrated numerically.
In contrast to the previous cases, condensation is not found for increasing pump rate.
It is rather obtained for decreasing pump rate, which implies decondensation with increasing pump rate.

\section{Absorption and Emission Spectra}
\label{supp:spectra}

All parameter values used for simulations match current photon BEC experiments \cite{2017arXiv170706789D}. The only difference between simulations and typical experiment values is in the mode spacing.
We use a mode spacing of $30 \trads$ ({\it i.e.} $30/(2\pi) 10^{12}\mbox{Hz}$) as this enables us to consider a system with few occupied modes, keeping it computationally tractable.
Spacings in current experiments are a factor of $3$ smaller, but there is no fundamental difficulty in experiments with larger spacings.

\begin{figure}[htpb]
	\centering
	\includegraphics[width=\columnwidth]{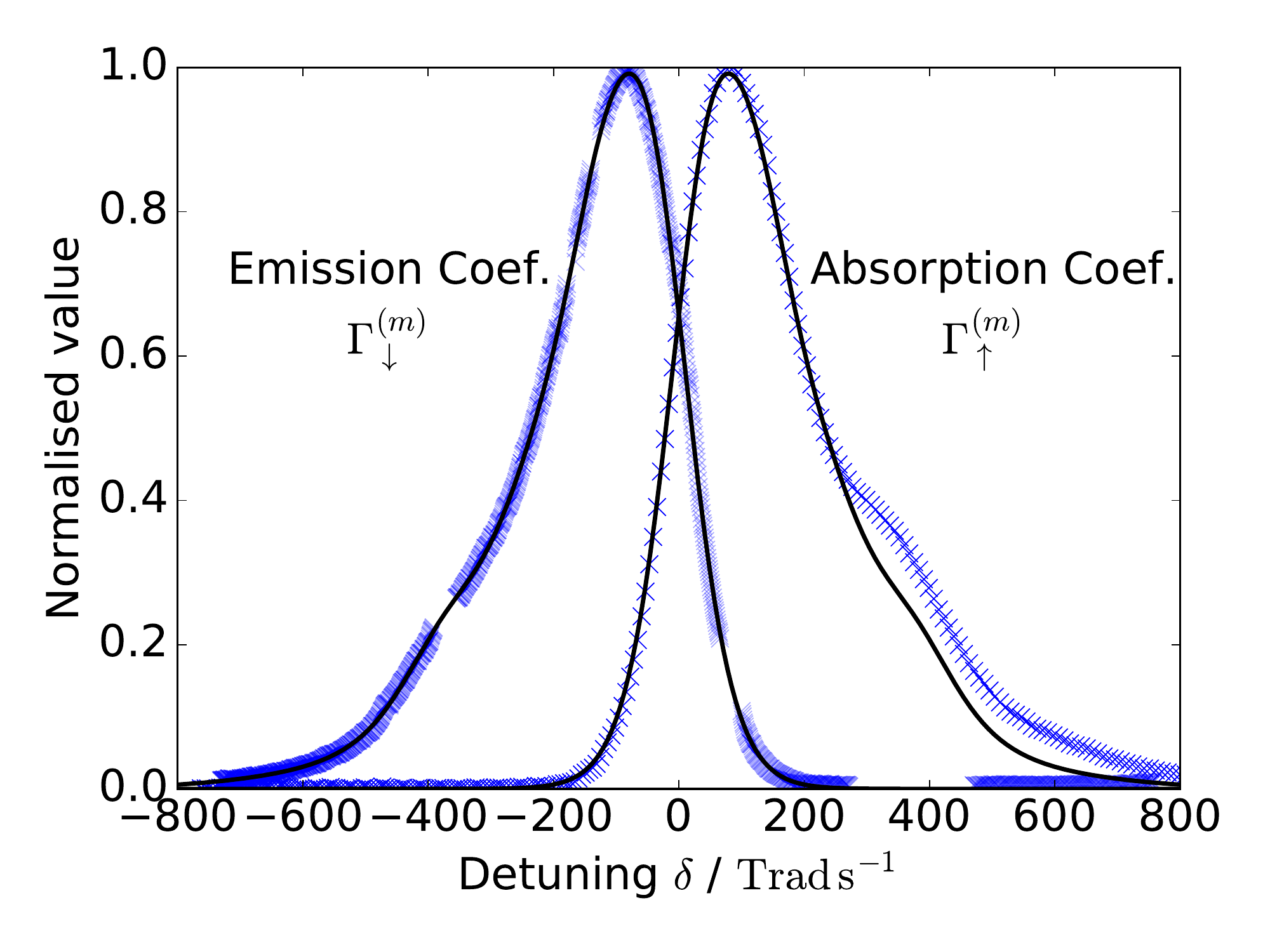}
	\caption{(Crosses) The experimental absorption and emission data. (Solid) The fits to this data.}
	\label{fig:ab_em}
\end{figure}

The absorption (emission) of light from (into) a cavity mode with detuning $\delta$ by molecules is derived from experimental data~\cite{nyman_robert_andrew_2017_569817}, shown in Fig.~\ref{fig:ab_em}.
The blue crosses indicate experimental data with a peak value of $1.2/s$.
In our simulations we use  $1.2\times 10^{-9}\kappa$ as peak value,
and we use the fitted dependence satisfying the Kennard-Stepanov/McCumber relation, depicted in black.

The cavity modes have detunings of
$\delta = \delta_{[0,0]} + m_x \omega_x + m_y \omega_y$
with $\omega_x = 30 \trads$ and $\omega_y = 30.3 \trads$, and $m_x$ and $m_y$ are the integers labeling each mode;
$\delta_{[0,0]}$ is the detuning of the lowest mode from the molecular splitting and ranges from $-350 \trads$ to $-180 \trads$.

\section{Threshold}
\label{supp:threshold}

The threshold condition used in the simulations is an increase in the population of a mode greater than 3 orders of magnitude over an increase in pump rate of 10\%. Here we justify our claim that the precise values used do not significantly change the phase diagram by presenting the large difference in mode population between the condensed and un-condensed modes.

\begin{figure}[t]
	\centering
	\includegraphics[width=\columnwidth]{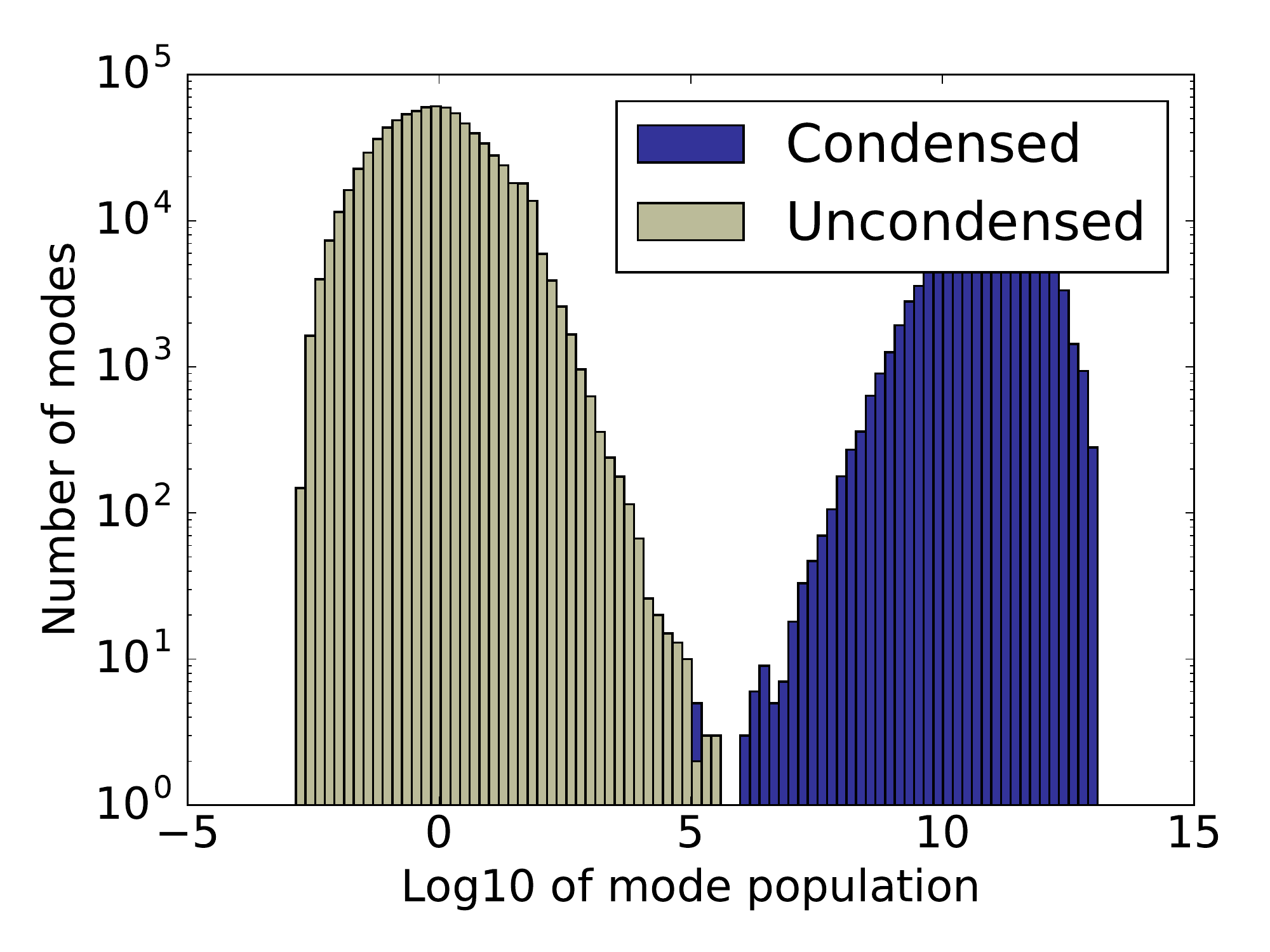}
	\caption{Histogram of the mode populations for all modes and all parameter values depicted in the phase diagram, Fig.~\ref{fig2}.}
	\label{fig:hist}
\end{figure}

Fig.~\ref{fig:hist} depicts a histogram for the occurrence of given mode populations.
The data depicted in green corresponds to all modes considered un-condensed in the phase diagram depicted in Fig.~\ref{fig2},
and data depicted in blue corresponds to modes categorized as condensed.
As one can see, each condensed mode has a population larger than $10^{5.8}$ and each un-condensed mode has a population smaller than $10^{5.8}$,
with the exception of 8 mis-categorized points with a population of about $10^5$.
Inspection of these points reveals that they are part way through a condensation or de-condensation, and therefore constitute the shifting of a phase boundary by only a single sample.
Changing the threshold condition to a change in population of $10^{3.4}$ instead of $10^{3}$ would 
result in perfect agreement between the threshold condition and the bi-modal distribution depicted in Fig.~\ref{fig:hist}.

\section{Dependence on pump spot size}
\label{supp:size}
A natural parameter which is easily varied in an experiment is the width of the pump profile. In the main paper, we consider an essentially homogeneous pump profile.
Narrow pump profiles have already been used to experimentally achieve multimode condensation,
and they do so at lower total pump rates \cite{Marelic2016} since a larger fraction of the pump power is concentrated on the central, condensed modes.
We therefore also investigated the dependence of the photon gas on the pump width,
with a Gaussian pump profile centered around the center of the cavity.

If the width of the pump profile is narrower than the spatial extent of the lowest cavity mode, then pumping is mostly restricted to this mode. The majority of the pumped dye molecules are therefore clamped by this mode, which prevents multimode condensation. One therefore obtains single-mode BEC even for very strong pumping.

With increasing pump width excited modes also experience pumping, and sufficiently strong pumping results in multimode condensation. As the pump spot size is increased further, the pump power is distributed over more modes, and some modes lose their macroscopic occupation. There is thus an optimal pump spot size at which multi- mode condensation can be achieved with minimal pump rate. We found this size to be about $1.4$ times larger than the mean spatial extent of the ground-state wavefunction.

\bibliographystyle{prsty}
\bibliography{citations-1}

\end{document}